\documentstyle[PASJadd]{PASJ95}
\input epsf.sty

\begin{document}
\title{OHS: OH-Airglow Suppressor for the Subaru Telescope}
\author{
Fumihide {\sc Iwamuro},$^1$
Kentaro {\sc Motohara},$^2$
Toshinori {\sc Maihara},$^3$\\
Ryuji {\sc Hata},$^1$
and
Takashi {\sc Harashima},$^1$
\\ [12pt]
$^1$ {\it Department of Physics, Kyoto University, Kitashirakawa, Kyoto 606-8502}\\
{\it iwamuro@cr.scphys.kyoto-u.ac.jp}\\
$^2${\it Subaru Telescope, National Astronomical Observatory, 650 North Aohoku Place, Hilo, HI 96720, USA}\\
$^3${\it Department of Astronomy, Kyoto University, Kitashirakawa, Kyoto 606-8502}\\
}
\abst{
This paper describes an OH-airglow Suppressor (OHS) for the infrared Nasmyth focus 
of the Subaru telescope. OHS has the capability of eliminating 224 airglow-lines 
in the $J$- and $H$-bands, which are major sources of background radiation 
at near-infrared wavelengths up to 2 $\mu$m. Specifically, it is a pre-optics system
installed between the telescope and an infrared camera/spectrograph (CISCO).
The suppressor reduces sky background emissions to 1/25 and its throughput is 40\%.
As a result, the S/N gain achieved with OHS is more than 1 mag compared to the 
typical spectroscopic approach. The limiting magnitude measured during a test 
observing run was found to be $H$ = 21.1 mag ($\lambda/\Delta\lambda$ = 210, S/N = 5) in 
the standard 4000 s exposure sequence.}

\kword{cosmology: observations --- instrumentation: spectrographs --- infrared: general}

\maketitle
\thispagestyle{headings}

\section
{Introduction}
Rapid progress has continued to be made in the study of the high-redshift universe over the 
last decade by the development of telescopes, large-format arrays, and observation techniques. 
To date, a considerable number of high-$z$ galaxies have been spectroscopically identified, 
and the nature of these galaxies has been investigated, mainly using their broad-band colors. 
However, it is impossible to separate the effects of age and extinction only from these colors. 
This problem has prevented the exact features of high-$z$ objects from being obtained. 
More detailed information --- rest-frame optical spectra --- is required for further studying 
these objects.

Although it is extremely difficult to obtain the rest-frame optical spectra of distant galaxies, 
several groups have succeeded in detecting some emission lines of high-$z$ radio
galaxies (Eales, Rawlings 1993, 1996; Evans 1998) and of Ly break galaxies (Pettini et al. 1998)
in the near-infrared $J$- to $K$-bands. Two antithetical concepts can be seen in their
results. One is a lower spectral resolution to detect a faint continuum and to achieve a wider 
wavelength coverage; the other is a higher spectral resolution to avoid the effects of 
strong airglow-lines. OHS is an instrument that can merge these two concepts, 
achieving a high sensitivity for a faint continuum and a wide spectral coverage. 

Before designing OHS for the Subaru telescope, we worked forwards confirming the basic performance 
of the suppressor by using a proto-type instrument (Iwamuro et al. 1994) from 1992 to 1995 in the 
University of Hawaii's 2.2 m telescope on Mauna Kea. The total design of OHS described here is 
based on our experience from these test observations. After that, we developed an infrared 
camera/spectrograph (CISCO: Motohara et al. 1998) as the back-end instrument of OHS, and OHS 
itself was mounted on the Nasmyth stage of the Subaru telescope at the end of 1999.

The optical design and mechanical design of OHS are described in sections 2 and 3, the details 
of the airglow mask are reported in section 4, and the performance of the OHS/CISCO system during 
its first nine months of operation is summarized in section 5.

\section
{Optical Design}
Figure 1 shows the optical layout of OHS. The focal ratio of the infrared Nasmyth
focus of the Subaru telescope is $f$/13.2 with the image rotator. As the figure shows, 
incident rays coming through the entrance slit in the gap of the two gratings are reflected 
by the collimator mirror, upper grating, camera mirror, mask mirror, camera mirror again, 
lower grating, refocus mirror, and pick-off mirror. Here, the incident beam, including 
the sky background and the object flux, is dispersed by the upper grating and is focused 
onto the mask mirror, where a bar-code-like mask is installed. This mask absorbs almost 
all airglow-lines; the remaining parts are reflected and re-combined by the lower grating. 
The upper layer (from the entrance slit to the mask mirror) and the lower layer (from the 
mask mirror to the pick-off mirror) must be separated to pick out the refocused slit image 
from the optical path between the refocus mirror and the entrance slit without attenuation. 

This configuration also reduces the contribution of scattered light from bright 
incident beam before airglow rejection to the final image. Any optical aberration 
caused by the spherical camera mirror is reduced by the collimator mirror and the refocus 
mirror with their hyperbolic surfaces. However, the separation between the upper and lower 
layers is still restricted, i.e., the size of the entrance slit is 15 mm (28$''$) long 
by 0.5 mm (0$.\hspace{-2pt}''$93) or 0.25 mm (0$.\hspace{-2pt}''$47) wide (see section 3).

The upper and lower gratings have exactly the same optical parameters: 245 lines mm$^{-1}$
and a blaze angle of 37$^{\circ}$. These gratings disperse and then re-combine $J$-band 
(1.11--1.35 $\mu$m) and $H$-band (1.48--1.80 $\mu$m) light with the order of fourth and third,
respectively. The dispersed $J$- and $H$-band light is focused onto the mask mirror 
simultaneously by a large spherical camera mirror 1.3 m in diameter. The spectral resolution 
on the mask mirror is $\lambda/\Delta\lambda$ = 5500 for a 0$.\hspace{-2pt}''$93 slit 
(and double for a narrower slit), while the rejection resolution is 3900, because the 
element of the mask is 1.4-times wider than the slit image (see section 4). 
Figure 2 shows spot diagrams at the end of the OHS optics and on the mask mirror.
Since the slit length is quite small compared with the focal length of the collimator,
the spot size is almost fixed inside the slit. The aspheric corrector plates 
placed in front of the gratings can reduce the spot size noticeably, but can also cause 
a lower throughput and ghost airglow images on the mask mirror, posing a serious problem 
for the OHS concept. The squares in the spot diagrams correspond to the pixel scale 
of CISCO (0$.\hspace{-2pt}''$105); the optical aberration in CISCO (Motohara et al. 1998) 
is slightly larger than these squares. Accordingly, the optical aberration in OHS 
does not have a large effect on the final image.

\section
{Mechanical Design}
All of the optical elements of OHS are equipped in the upper frame supported by three
hydraulic jacks on the base frame, which allows the direction of the instrument to be adjusted
(figure 3). The base frame is also slidable to the waiting position by hand, despite a total weight of 
3000 kg and a size of 5m $\times$ 3m. The entrance slit, the gratings, and the pickoff mirror supports are mounted on 
the retractable stage, which is drawn back at the waiting position to avoid any collision
with the structure of the telescope. The entrance slit has five positions: pinhole, 
narrower (0$.\hspace{-2pt}''$47 $\times$ 28$''$) slit, wider (0$.\hspace{-2pt}''$93 $\times$ 28$''$) slit, 
open (20$''\times$28$''$), and close. The collimator/refocus mirror support and mask 
mirror support are mounted on the adjustable stages driven by stepper motors along the 
optical axis. The dispersion power on the mask mirror can be tuned to the scale of the 
airglow mask by adjusting the position of the mask mirror along the optical axis, 
because the separation between dispersed beams changes as the position is moved (see figure 1). 
The airglow mask focus is adjusted by moving the position of the collimator mirror. Since 
these adjustments can be done only for real airglow-lines, the positions of these elements
are controlled from the observation room.

CISCO is supported by an adjustable stage featuring three electric jacks and three
sliding or rotating stages (figure 4). CISCO is also used as a general infrared camera
with a field of view of 1$.\hspace{-2pt}'$8 square attached to the Nasmyth focus directly 
with this adjustable stage (after OHS is moved to the waiting position). We can adjust 
the direction of the optical axis and the position of the camera, whichever the stage is 
placed.

\section
{Airglow Mask}
The airglow mask is made of a thin stainless-steel plate 0.2 mm thick, and processed by 
photochemical etching (figure 5). The surface of the mask is coated with special black paint, 
which absorbs more than 99.9\% of all near-infrared light up to 2 $\mu$m. A total of 
224 OH and O$_2$ airglow-lines in the $J$- and $H$-bands are rejected here, while 
the effective opening area is 74.4\%. The width of the element of the mask is 0.7 mm, 
which is a factor 1.4 times the width of the slit image of 0.5 mm (this is because there 
may be small differences between real airglow-lines and the position of the mask element). 
The mask has overlapping patterns of airglow-lines in the $J$- and $H$-bands 
with different orders of the grating. Although this configuration is not favorable 
considering the instrumental throughput, we can obtain airglow-suppressed spectra in 
both bands simultaneously. From this point of view, we do not reject all broad molecular 
oxygen lines near 1.28 $\mu$m; this rejection requires a very wide mask for complete removal.

The wavelength of the OH-airglow emission lines was calculated from the transition 
levels of the OH radical reported by Coxon (1980), and Coxon, Foster (1982). The 
judgment of whether to reject the lines or not was made by referring to the airglow 
spectra observed in Maihara et al. (1993). The airglow lines rejected by this airglow 
mask are listed in table 1.

\section
{Test Observations and Verified Performance}
Following the installation of OHS at the end of 1999, test observation runs were 
carried out about every two months. In this section, we first summarize the results 
of these observations, and then describe the performance of this system, 
verified by several spectra of faint objects.

The typical observation procedure starts from what we call the ``imaging mode'', 
in which the slit of OHS is opened and CISCO is operated as an infrared camera (figure 6). 
The airglow-suppressed area corresponds to the black stripe at the center of the sky
image, called the ``dark lane'', where the airglow-lines are in phase with the airglow 
mask. The effective area of the dark lane is 1$.\hspace{-2pt}''$3 $\times$ 28$''$.
After an object is guided into this area, about a dozen frames are taken by nodding 
the telescope along the dark lane between the exposure sequences. These frames are
used not only for photometry, but also to confirm the object location in the 
dark lane at each nodding position. Next, the observation mode is switched to what is 
called the ``spectroscopic mode'', in which the slit of OHS is moved to the center 
of the dark lane; in CISCO, the width of the variable slit is reduced 
and the $JH$-band grism is selected. A standard exposure time of 1000 s reaches 
the background-limited condition with multiple-readouts of six-times the detector.
At least four exposures should be made to avoid bad pixels and unexpected 
hot pixels; accordingly, the standard exposure time comes to be more than 4000 s.

The total system efficiency was measured through observations of the UKIRT 
faint-standard stars in the imaging mode. The efficiency was 8.3\% in the $J$-band 
and 12.2\% in the $H$-band. The throughput of the OHS part was estimated to be 40\% 
in both bands by a comparison of the throughput with and without OHS.

The spectroscopic images taken by OHS are compared with images taken by the direct 
spectroscopy of CISCO in figure 7. The observable spectral range for the CISCO 
$JH$-band spectroscopy is from 1.044 $\mu$m to 1.816 $\mu$m, which is limited by the order 
sorting filter paired with the grism, while the range becomes discontinuous 
with OHS, because the light out of either end of the mask mirror does not
reach the detector. Almost all airglow-lines, except for broad O$_2$ lines in the $J$-band, 
are rejected by OHS; this is also shown in figure 8. The rejection factor of the 
airglow-lines is 25. 

The gain of OHS is illustrated in the lower half of figure 7. The object is radio 
galaxy 4C +40.36 ($z$ = 2.269) as observed by CISCO (May 24) and by CISCO with OHS
(May 23). It is quite obvious that OHS has a large advantage in terms of sensitivity 
for faint objects, even with a shorter exposure time. Since the object and background 
flux are in proportion to $\eta$ and $\eta /f_{\rm s}$, respectively ($\eta$: throughput, 
$f_{\rm s}$: suppression factor), the gain of OHS is estimated simply by\\

\begin{equation}
Gain = \frac{\eta}{\sqrt{\frac{\eta}{f_{\rm s}}}} = \sqrt{\eta f_{\rm s}} .
\end{equation}

The estimated gain becomes 3.16 or 1.25 mag from the measured values of $\eta = 0.4$ 
and $f_{\rm s} = 25$. The contribution of the read noise to the total background noise is 
about 10\% in the case of 1000 s exposure; accordingly, the actual gain, including 
the read noise, is 2.87 or 1.15 mag. The limiting magnitude, as confirmed by test 
observations, is $H$ = 21.1 mag ($\lambda/\Delta\lambda$ = 210, S/N = 5) in the standard 
4000 s exposure sequence with a 0$.\hspace{-2pt}''$93 slit. 

Figure 9 shows spectroscopic images of several faint objects taken with OHS during 
the last test observation run. In these images, the noisy regions between the bands 
contain no data. This is because OHS throughput corrections are applied using the 
spectroscopic standard stars observed just after each object. 
Table 2 summarizes the specifications and performance. Actually, many radio galaxies 
and quasars have already been observed by the OHS/CISCO system, and excellent performance 
levels have been exhibited, not only for emission lines, but also for continuum features, 
through the observations.

\par
\vspace{1pc}\par
The present results were accomplished during a test observation run of the Subaru 
telescope. We are therefore indebted to all members of the Subaru Observatory, 
NAOJ, Japan. We would like to express our thanks to the engineering staff of Mitsubishi 
Electric Co. for their fine operation of the telescope. This work was also supported by 
a Grant-in-Aid for Scientific Research (B), Japan (No. 11440065).

\section*{References}
\re Coxon, J. A.\ 1980, Can.\ J.\ Phys.,\ 58, 933
\re Coxon, J. A., \& Foster S. C.\ 1982, Can.\ J.\ Phys.,\ 60, 41
\re Eales, S. A., \& Rawlings, S.\ 1993, ApJ, 411, 67
\re Eales, S. A., \& Rawlings, S.\ 1996, ApJ, 460, 68
\re Evans, A. S.\ 1998, ApJ, 498, 553
\re Iwamuro, F., Maihara, T., Oya, S., Tsukamoto, H., Hall, D. N. B., Cowie, L. L., Tokunaga, A. T., \& Pickles, A. J.\ 1994, PASJ, 46, 515
\re Maihara, T., Iwamuro, F., Yamashita, T., Hall, D. N. B., Cowie, L. L., Tokunaga, A. T., \& Pickles, A.\ 1993, PASP, 105, 940
\re Motohara, K., Maihara, T., Iwamuro, F., Oya, S., Imanishi, M., Terada, H., Goto, M., Iwai, J.\ et al.\ 1998, Proc.\ SPIE, 3354, 659
\re Pettini, M., Kellogg, M., Steidel, C. C., Dickinson, M., Adelberger, K. L., \& Giavalisco, M.\ 1998, ApJ, 508, 539
\onecolumn
\begin{table*}[t]
\small
\begin{center}
Table~1.\hspace{4pt}{List of targeted airglow-lines.}\\
\end{center}
\vspace{6pt}
\begin{tabular*}{\textwidth}{@{\hspace{\tabcolsep}
\extracolsep{\fill}}llllllll}
\hline\hline\\ [-6pt]
\multicolumn{1}{c}{Line}& \multicolumn{1}{c}{$\lambda$ ($\mu$m)}&\multicolumn{1}{c}{Line}&\multicolumn{1}{c}{$\lambda$ ($\mu$m)}&\multicolumn{1}{c}{Line}&\multicolumn{1}{c}{$\lambda$ ($\mu$m)}&\multicolumn{1}{c}{Line}&\multicolumn{1}{c}{$\lambda$ ($\mu$m)}\\
[4pt]\hline\\[-6pt]
OH(3--1)R1(1) &   1.48877 & OH(5--3)Q1(2) &   1.67088 & OH(5--3)P2(6) &   1.73511 & OH(10--7)Q1(3) &   1.49850\\
OH(3--1)R2(1) &   1.49319 & OH(5--3)Q2(2) &   1.67026 & OH(5--3)P1(7) &   1.75294 & OH(5--2)P1(5) &   1.10900\\
OH(3--1)R1(2) &   1.48331 & OH(5--3)Q1(3) &   1.67325 & OH(5--3)P2(7) &   1.75012 & OH(5--2)P2(5) &   1.10724\\
OH(3--1)R2(2) &   1.48644 & OH(5--3)Q2(3) &   1.67247 & OH(5--3)P1(8) &   1.76855 & OH(5--2)P1(6) &   1.11560\\
OH(3--1)R1(3) &   1.47837 & OH(5--3)Q1(4) &   1.67636 & OH(6--4)P1(2) &   1.78802 & OH(5--2)P2(6) &   1.11409\\
OH(3--1)R2(3) &   1.48058 & OH(5--3)Q1(5) &   1.68024 & OH(6--4)P2(2) &   1.78115 & OH(5--2)P1(7) &   1.12279\\
OH(4--2)R1(1) &   1.56549 & OH(6--4)Q1(1) &   1.76532 & OH(6--4)P1(3) &   1.79939 & OH(6--3)P1(2) &   1.15388\\
OH(4--2)R2(1) &   1.57025 & OH(6--4)Q2(1) &   1.76498 & OH(6--4)P2(3) &   1.79347 & OH(6--3)P1(3) &   1.15917\\
OH(4--2)R1(2) &   1.55977 & OH(6--4)Q1(2) &   1.76718 & OH(6--3)R1(1) &   1.13542 & OH(6--3)P2(3) &   1.15652\\
OH(4--2)R2(2) &   1.56316 & OH(6--4)Q2(2) &   1.76649 & OH(6--3)R2(1) &   1.13771 & OH(6--3)P1(4) &   1.16507\\
OH(4--2)R1(3) &   1.55461 & OH(6--4)Q1(3) &   1.76984 & OH(6--3)R1(2) &   1.13312 & OH(6--3)P2(4) &   1.16278\\
OH(4--2)R2(3) &   1.55702 & OH(6--4)Q2(3) &   1.76896 & OH(6--3)R2(2) &   1.13469 & OH(6--3)P1(5) &   1.17161\\
OH(4--2)R1(4) &   1.55009 & OH(6--4)Q1(4) &   1.77334 & OH(6--3)R1(3) &   1.13128 & OH(6--3)P2(5) &   1.16963\\
OH(4--2)R2(4) &   1.55179 & OH(6--4)Q1(5) &   1.77771 & OH(6--3)R2(3) &   1.13234 & OH(6--3)P1(6) &   1.17880\\
OH(4--2)R1(5) &   1.54621 & OH(2--0)P1(5) &   1.47998 & OH(6--3)R1(4) &   1.12994 & OH(6--3)P2(6) &   1.17708\\
OH(4--2)R2(5) &   1.54742 & OH(2--0)P2(5) &   1.47724 & OH(6--3)R2(4) &   1.13063 & OH(6--3)P1(7) &   1.18665\\
OH(4--2)R1(6) &   1.54302 & OH(2--0)P1(6) &   1.49090 & OH(6--3)R1(5) &   1.12912 & OH(7--4)P1(2) &   1.22292\\
OH(5--3)R1(1) &   1.65024 & OH(2--0)P2(6) &   1.48862 & OH(7--4)R1(1) &   1.20308 & OH(7--4)P2(2) &   1.21964\\
OH(5--3)R2(1) &   1.65538 & OH(2--0)P1(7) &   1.50261 & OH(7--4)R2(1) &   1.20559 & OH(7--4)P1(3) &   1.22869\\
OH(5--3)R1(2) &   1.64421 & OH(2--0)P2(7) &   1.50067 & OH(7--4)R1(2) &   1.20070 & OH(7--4)P2(3) &   1.22577\\
OH(5--3)R2(2) &   1.64790 & OH(2--0)P1(8) &   1.51510 & OH(7--4)R2(2) &   1.20242 & OH(7--4)P1(4) &   1.23515\\
OH(5--3)R1(3) &   1.63885 & OH(3--1)P1(2) &   1.52410 & OH(7--4)R1(3) &   1.19886 & OH(7--4)P2(4) &   1.23259\\
OH(5--3)R2(3) &   1.64147 & OH(3--1)P2(2) &   1.51871 & OH(7--4)R2(3) &   1.20001 & OH(7--4)P1(5) &   1.24233\\
OH(5--3)R1(4) &   1.63418 & OH(3--1)P1(3) &   1.53324 & OH(7--4)R1(4) &   1.19758 & OH(7--4)P2(5) &   1.24008\\
OH(5--3)R2(4) &   1.63604 & OH(3--1)P2(3) &   1.52878 & OH(7--4)R2(4) &   1.19833 & OH(7--4)P1(6) &   1.25024\\
OH(5--3)R1(5) &   1.63023 & OH(3--1)P1(4) &   1.54322 & OH(7--4)R1(5) &   1.19688 & OH(7--4)P2(6) &   1.24827\\
OH(5--3)R2(5) &   1.63155 & OH(3--1)P2(4) &   1.53953 & OH(8--5)R1(1) &   1.28070 & OH(7--4)P1(7) &   1.25890\\
OH(5--3)R1(6) &   1.62703 & OH(3--1)P1(5) &   1.55403 & OH(8--5)R2(1) &   1.28346 & OH(8--5)P1(2) &   1.30217\\
OH(6--4)R1(1) &   1.74499 & OH(3--1)P2(5) &   1.55098 & OH(8--5)R1(2) &   1.27826 & OH(8--5)P2(2) &   1.29857\\
OH(6--4)R2(1) &   1.75059 & OH(3--1)P1(6) &   1.56570 & OH(8--5)R2(2) &   1.28016 & OH(8--5)P1(3) &   1.30853\\
OH(6--4)R1(2) &   1.73867 & OH(3--1)P2(6) &   1.56313 & OH(8--5)R1(3) &   1.27644 & OH(8--5)P2(3) &   1.30528\\
OH(6--4)R2(2) &   1.74270 & OH(3--1)P1(7) &   1.57821 & OH(8--5)R2(3) &   1.27772 & OH(8--5)P1(4) &   1.31568\\
OH(6--4)R1(3) &   1.73308 & OH(3--1)P2(7) &   1.57603 & OH(8--5)R1(4) &   1.27528 & OH(8--5)P2(4) &   1.31279\\
OH(6--4)R2(3) &   1.73597 & OH(3--1)P1(8) &   1.59159 & OH(8--5)R2(4) &   1.27611 & OH(8--5)P1(5) &   1.32365\\
OH(6--4)R1(4) &   1.72829 & OH(4--2)P1(2) &   1.60309 & OH(8--5)R1(5) &   1.27480 & OH(8--5)P2(5) &   1.32109\\
OH(6--4)R2(4) &   1.73034 & OH(4--2)P2(2) &   1.59726 & OH(6--3)Q1(1) &   1.14399 & OH(8--5)P1(6) &   1.33247\\
OH(6--4)R1(5) &   1.72431 & OH(4--2)P1(3) &   1.61286 & OH(6--3)Q2(1) &   1.14378 & OH(8--5)P2(6) &   1.33020\\
OH(6--4)R2(5) &   1.72576 & OH(4--2)P2(3) &   1.60797 & OH(6--3)Q1(2) &   1.14516 & OH(8--5)P1(7) &   1.34216\\
OH(6--4)R1(6) &   1.72117 & OH(4--2)P1(4) &   1.62354 & OH(6--3)Q2(2) &   1.14472 & OH(8--5)P2(7) &   1.34013\\
OH(6--4)R2(6) &   1.72220 & OH(4--2)P2(4) &   1.61946 & OH(6--3)Q1(3) &   1.14683 & OH(8--5)P1(8) &   1.35274\\
OH(6--4)R1(7) &   1.71891 & OH(4--2)P1(5) &   1.63513 & OH(6--3)Q1(4) &   1.14903 & O2            &   1.26850\\
OH(6--4)R1(8) &   1.71753 & OH(4--2)P2(5) &   1.63172 & OH(7--4)Q1(1) &   1.21226 & O2            &   1.26875\\
OH(3--1)Q1(1) &   1.50555 & OH(4--2)P1(6) &   1.64765 & OH(7--4)Q2(1) &   1.21204 & O2            &   1.26900\\
OH(3--1)Q2(1) &   1.50529 & OH(4--2)P2(6) &   1.64476 & OH(7--4)Q1(2) &   1.21359 & O2            &   1.26925\\
OH(3--1)Q1(2) &   1.50690 & OH(4--2)P1(7) &   1.66110 & OH(7--4)Q2(2) &   1.21311 & O2            &   1.26950\\
OH(3--1)Q2(2) &   1.50640 & OH(4--2)P2(7) &   1.65863 & OH(7--4)Q1(3) &   1.21549 & O2            &   1.26975\\
OH(3--1)Q1(3) &   1.50883 & OH(4--2)P1(8) &   1.67551 & OH(7--4)Q1(4) &   1.21799 & O2            &   1.27000\\
OH(3--1)Q1(4) &   1.51137 & OH(5--3)P1(2) &   1.69037 & OH(8--5)Q1(1) &   1.29057 & O2            &   1.27095\\
OH(4--2)Q1(1) &   1.58333 & OH(5--3)P2(2) &   1.68405 & OH(8--5)Q2(1) &   1.29036 & O2            &   1.27260\\
OH(4--2)Q2(1) &   1.58303 & OH(5--3)P1(3) &   1.70087 & OH(8--5)Q1(2) &   1.29212 & O2            &   1.27460\\
OH(4--2)Q1(2) &   1.58482 & OH(5--3)P2(3) &   1.69551 & OH(8--5)Q2(2) &   1.29159 & O2            &   1.27955\\
OH(4--2)Q2(2) &   1.58425 & OH(5--3)P1(4) &   1.71236 & OH(8--5)Q1(3) &   1.29432 & O2            &   1.28245\\
OH(4--2)Q1(3) &   1.58693 & OH(5--3)P2(4) &   1.70783 & OH(8--5)Q1(4) &   1.29722 & O2            &   1.28450\\
OH(4--2)Q1(4) &   1.58973 & OH(5--3)P1(5) &   1.72486 & OH(10--7)Q1(1) &   1.49302 & O2            &   1.28660\\
OH(5--3)Q1(1) &   1.66924 & OH(5--3)P2(5) &   1.72103 & OH(10--7)Q2(1) &   1.49294 & O2            &   1.58045\\
OH(5--3)Q2(1) &   1.66892 & OH(5--3)P1(6) &   1.73838 & OH(10--7)Q1(2) &   1.49528 & O2            &   1.58085\\
[4pt]\hline\\[-6pt]
\end{tabular*}
\end{table*}
\newpage
\begin{table*}[t]
\begin{center}
Table~2.\hspace{4pt}{Specifications and performance.}\\
\end{center}
\vspace{6pt}
\begin{tabular*}{\textwidth}{@{\hspace{\tabcolsep}
\extracolsep{\fill}}ll}
\hline\hline\\ [-6pt]
Wavelength coverage& 1.477--1.804 $\mu$m ($H$-band), 1.108--1.353 $\mu$m ($J$-band)\\
Slit size& 0$.\hspace{-2pt}''$93 or 0$.\hspace{-2pt}''$47 $\times 28''$\\
Throughput& 40\%\\
Suppression factor& 25\\
Gain by OHS& 2.87 (1.15 mag)\\
Total system efficiency& 12.2\% ($H$-band), 8.3\% ($J$-band)\\
Final spectral resolution& 210 or 420\\
Limiting magnitude& $H$ = 21.1 mag (S/N = 5, 4000 s)\\
[4pt]\hline\\[-6pt]
\end{tabular*}
\end{table*}

\newpage
\begin{figure}[p]
\caption{Optical design layout of OHS. The focal lengths of all of the concave mirrors
are 2000 mm, and the centers of curvature of the camera and the mask mirrors are at the 
same position as the center of the entrance slit.}
\end{figure}
\begin{figure}[p]
\caption{Spot diagrams at 1.48, 1.56, 1.64, 1.72, and 1.80 $\mu$m in the $H$-band. The diagrams 
in the $J$-band are exactly the same as those in the $H$-band because their differences are 
equivalent to the order of the gratings. The squares of 56 $\mu$m correspond to the pixel scale of CISCO 
(0$.\hspace{-2pt}''$105 at the Nasmyth focus). a) Refocused spot image at the end of the OHS part, 
where the entrance slit of CISCO is located. b) Spot images on the mask mirror. The width of 
the element of the mask is 700 $\mu$m (see section 4).}
\end{figure}
\begin{figure}[p]
\caption{Assembly drawing of OHS.}
\end{figure}
\begin{figure}[p]
\caption{Photograph of the OHS/CISCO system mounted on the Nasmyth focus of the Subaru telescope.
CISCO is positioned to the side of OHS, at the top left in this photograph.}
\end{figure}
\begin{figure}[p]
\epsfxsize=17cm
\hspace{1cm}
\epsfbox{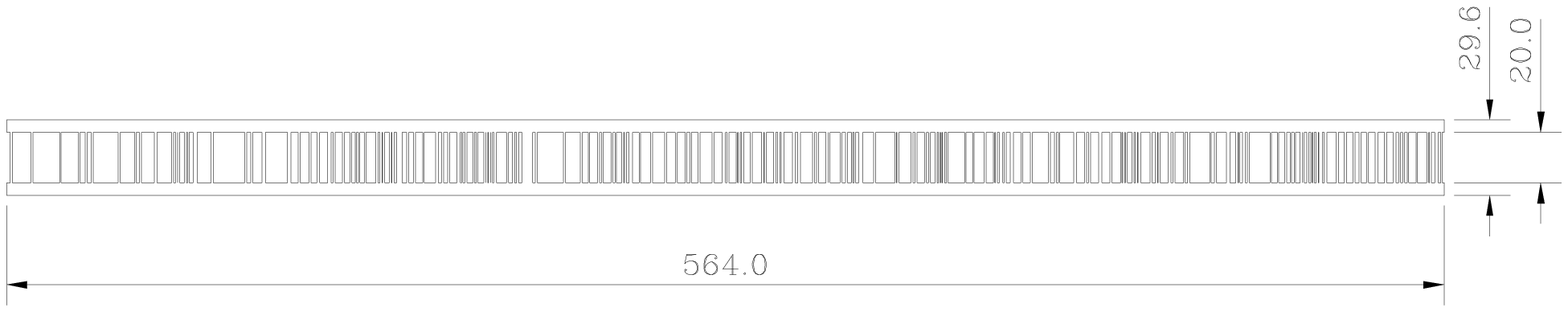}
\caption{Airglow mask made of a thin stainless-steel plate 0.2 mm thick. This mask has overlapping 
patterns of airglow-lines in the $J$- and $H$-bands, and the effective opening area is 74.4\%.}
\end{figure}
\begin{figure}[p]
\caption{Blank-sky image taken in the imaging mode. The field of view in this mode is
$20''\times28''$ and the airglow-suppressed area corresponds to the black stripe at the 
center of this image.}
\end{figure}
\begin{figure}[p]
\caption{Comparison between CISCO direct spectroscopy and the OHS/CISCO system.
a--c) Airglow spectra (a) taken by CISCO, (b) taken by the OHS/CISCO system, 
and (c) multiplied by 5. d--f) Reduced spectra of the radio galaxy 4C +40.36 ($z$ = 2.27)
(d) taken by CISCO with 1200 s exposure, (e) taken by the OHS/CISCO system with 
800 s exposure, and (f) taken by the OHS/CISCO system with 3200 s exposure. 
Five emission lines can be detected, i.e., [O {\sc ii}] 3727, [Ne {\sc iii}] 3869, 
H$\beta$, and [O {\sc iii}] 4959/5007.}
\end{figure}
\begin{figure}[p]
\epsfxsize=15cm
\hspace{1cm}
\epsfbox{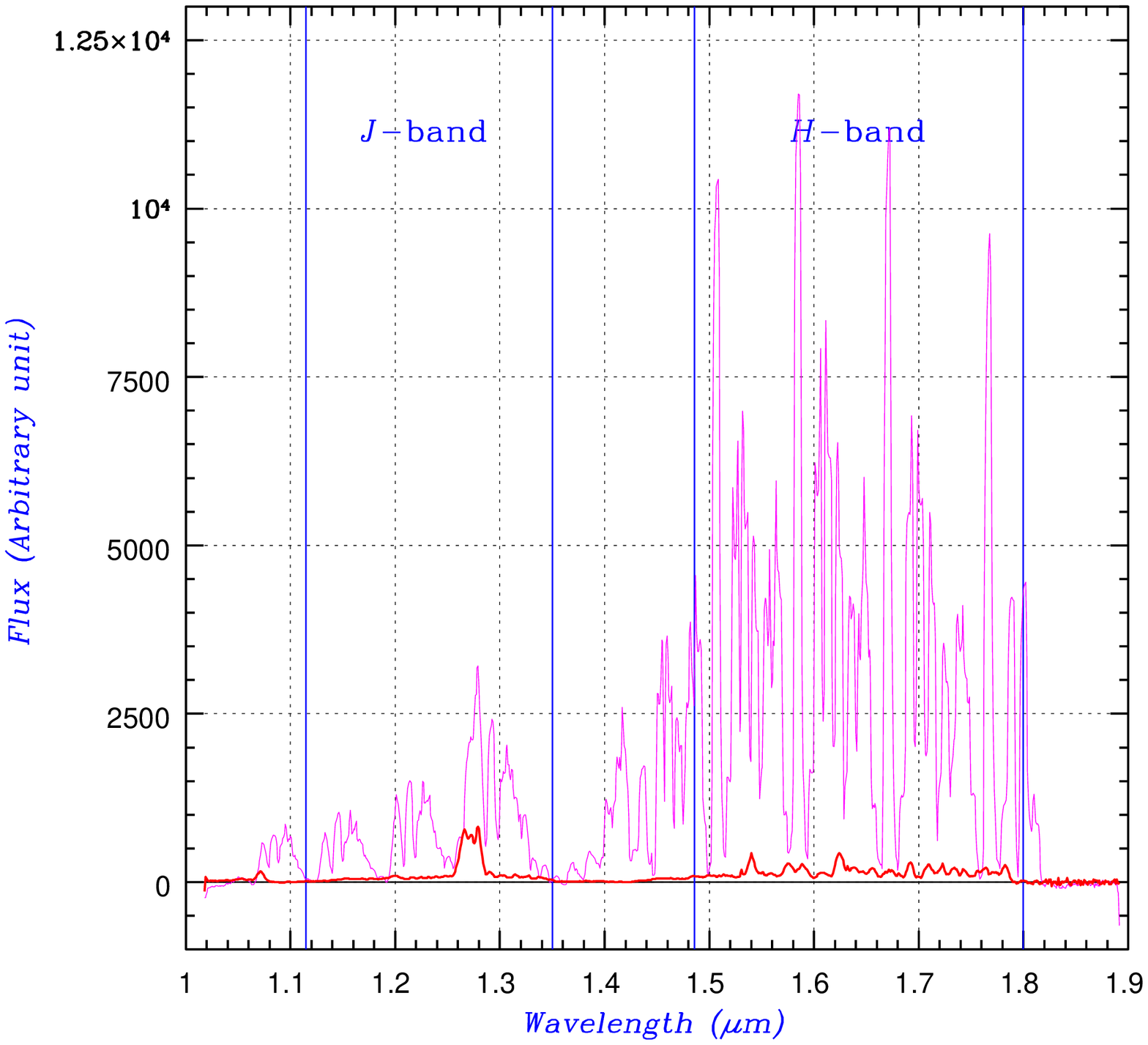}
\caption{Airglow spectra with and without OHS. The throughput of OHS (40\%) is 
corrected and the rejection factor of the airglow-lines is 30 except the region of 
the broad O$_2$ lines at $\sim 1.28 \mu$m. The vertical lines indicate the observable 
spectral range with OHS.}
\end{figure}
\begin{figure}[p]
\caption{$H$-band images (left) and $JH$-band spectra (right) of several faint objects.
a) Radio galaxy MRC 0156$-$252 ($H$ = 18.45, $z$ = 2.025, 4000 s) with a spectacular ``[O {\sc iii}] jet''.
b) SDSSp J021102.72$-$000910.3 ($H$ = 18.75, $z$ = 4.895, 6000 s) with a very broad Mg {\sc ii} emission. 
c) Radio galaxy MRC 0406$-$244 SE ($H$ = 19.11, $z$ = 2.428, 6000 s) with a knot structure.
d) An ERO candidate ($H$ = 19.75, $z$ = unknown, 6000 s).
e) BTM98 cK39W/E($H$ = 20.89, $z$ = 2.434, 5000 s) with [O {\sc ii}] and [O {\sc iii}] emission lines.
f) MLG93 G2($H$ = 22.04, $z$ = 3.429, 8000 s) with strong [O {\sc ii}] and [Ne {\sc iii}] lines.}
\end{figure}
\end{document}